\documentclass[preprint,12pt]{elsarticle}

\usepackage{amssymb}
\usepackage{amsmath}
\usepackage{hyperref}
\begin{document}

\begin{frontmatter}

\title{SST-DUNet: Automated preclinical functional MRI skull stripping using Smart Swin Transformer and Dense UNet} 

\author[label1]{Sima Soltanpour} 
\author[label2]{Rachel Utama}
\author[label2]{Arnold Chang}
\author[label3]{Md Taufiq Nasseef}
\author[label4,label5]{Dan Madularu}
\author[label2]{Praveen Kulkarni}
\author[label2]{Craig Ferris}
\author[label1]{Chris Joslin}

\affiliation[label1]{organization={School of Information Technology, Carleton University},
            addressline={1125 Colonel By Dr}, 
            city={Ottawa},
            postcode={K1S 5B6}, 
            state={ON},
            country={Canada}
}
\affiliation[label2]{organization={Center for Translational NeuroImaging (CTNI), Northeastern University},
            addressline={360 Huntington Ave}, 
            city={Boston},
            postcode={02115}, 
            state={MA},
            country={USA}
}
\affiliation[label3]{organization={Department of Mathematics, College of Science and Humanity Studies, Prince Sattam Bin Abdulaziz University},
            addressline={Al Kharj}, 
            city={Riyadh},
            country={Saudi Arabia}
} 
\affiliation[label4]{organization={Department of Psychology, Carleton University},
            addressline={1125 Colonel By Drive}, 
            city={Ottawa},
            postcode={K1S 5B6}, 
            state={Ontario},
            country={Canada}         
}
\affiliation[label5]{organization={Tessellis Ltd.},
            addressline={350 Legget Drive}, 
            city={Ottawa},
            postcode={K2K 0G7}, 
            state={Ontario},
            country={Canada}
}         

\begin{abstract}
Skull stripping is a common preprocessing step that is often performed manually in Magnetic Resonance Imaging (MRI) pipelines, including functional MRI (fMRI). This manual process is time-consuming and operator dependent. Automating this process is challenging for preclinical data due to variations in brain geometry, resolution, and tissue contrast. While existing methods for MRI skull stripping exist, they often struggle with the low resolution and varying slice sizes in preclinical fMRI data.
This study proposes a novel method called SST-DUNet, that integrates a dense UNet-based architecture with a feature extractor based on Smart Swin Transformer (SST) for fMRI skull stripping. The Smart Shifted Window Multi-Head Self-Attention (SSW-MSA) module in SST is adapted to replace the mask-based module in the Swin Transformer (ST), enabling the learning of distinct channel-wise features while focusing on relevant dependencies within brain structures. This modification allows the model to better handle the complexities of fMRI skull stripping, such as low resolution and variable slice sizes. To address the issue of class imbalance in preclinical data, a combined loss function using Focal and Dice loss is utilized. 
The model was trained on rat fMRI images and evaluated across three in-house datasets with a Dice similarity score of 98.65\%, 97.86\%, and 98.04\%.
The fMRI results obtained through automatic skull stripping using the SST-DUNet model closely align with those from manual skull stripping for both seed-based and independent component analyses. These results indicate that the SST-DUNet can effectively substitute manual brain extraction in rat fMRI analysis.
\end{abstract}


\end{frontmatter}

\section{Introduction}
\label{sec1}
Preclinical fMRI analysis plays a key role in understanding brain function and neurological disorders in animal models, with the goal of improving human clinical research \cite{zerbi2022use}. By studying functional connectivity, brain activity patterns, and neural dynamics, preclinical fMRI enables researchers to investigate the effects of diseases, drugs, and interventions in a controlled environment \cite{grandjean2023consensus}. 
However, the accuracy and reliability of fMRI analysis depend heavily on the preprocessing pipeline, which is essential for preparing data for preclinical analysis. 
3D Skull stripping is one of the crucial preprocessing steps, by extracting the brain image from a 3D MRI image of the head. The output of this process is a 3D binary mask that is used to distinguish a contiguous brain from the rest of the head. 
Manually extracting the brain is considered a common method for skull stripping, involving experts who manually outline the brain image from the rest of the head \cite{hsu2020automatic}. Although a manual approach can produce robust results, it is subjective, labor-intensive, and time-consuming \cite{feo2019towards}. To address this issue, semi or fully automated methods have been developed for skull stripping. Developing such algorithms not only enhances data quality but also improves the reproducibility and consistency of findings, driving advancements in neuroscience research.

In clinical MRI research related to the human brain, different automatic skull stripping algorithms have been proposed and widely applied \cite{jiang2021brain, pei2022general, hoopes2022synthstrip, moazami2023probabilistic, nishimaki2023pcss}. These methods are divided into two categories conventional and learning-based methods \cite{fatima2020state}.
However, these methods are not applicable to preclinical applications due to differences in brain and scalp geometry, image resolution, and image contrast. 
 
For preclinical skull stripping, the most prominent conventional algorithms include Rapid Automatic Tissue Segmentation (RATS) proposed by Oguz et al. \cite{oguz2014rats}, a template-derived approach proposed by Lohmeier et al. \cite{lohmeier2019atlasbrex}, SHape descriptor selected Extremal Regions after Morphologically filtering (SHERM) proposed by Liu et al. \cite{liu2020automatic}, and antsBrainExtraction (antsBE) proposed by MacNicol et al. \cite{macnicol2021atlas}.   
A common constraint observed in the above mentioned skull stripping methods is their performance variability based on factors such as brain size, shape, texture, and contrast. 
Recently, an atlas-based Veterinary Images Brain Extraction (VIBE) algorithm to extract brains on multi-contrast animal MRI has been proposed by Eddin et al. \cite{nour2023automatic}. However, this method has been tested on cat and dog brain MRI data and has not been evaluated on other animal brain datasets like rats and mice.

Progress in the field of machine learning has introduced potential automated approaches for preclinical skull stripping. Roy et al. \cite{roy2018deep} proposed a Convolutional Neural Netwrk (CNN) architecture with modified Google Inception \cite{szegedy2015going} using multiple atlases for 2D rodent brain MRI images. With the advancement of deep learning, more advanced architectures of CNN have been introduced for semantic segmentation. Among them, UNet, a widely used deep CNN model, has demonstrated remarkable effectiveness in this task. 2D UNet is utilized by Thai et al. \cite{thai2019using} for mouse skull stripping on diffusion weighted images. Hsu et al. \cite{hsu2020automatic} proposed a convolutional deep learning-based method for skull stripping of rat and mouse brains using UNet for 2D MRI data. 
The MRI image patches were randomly cropped to create the brain mask, which served as input for the 2D UNet architecture. The model's performance was assessed by comparing a manually traced brain mask by an anatomical expert with the one generated by the model. Deo Feo et al. \cite{de2021automated} introduced a multitask UNet (MUNet) designed for skull stripping and region segmentation. The training and validation of the model utilized anatomical MRI images from mice. 
Recently, Liang et al. \cite{liang2023automatic} proposed a skull stripping algorithm using U$^2$-Net for rat brain MRI. RUNet is applied by Chang et al. \cite{chang2023ru} for skull stripping of rat MRI after ischemic stroke. Also, Porter et al. \cite{porter2024fully} applied 2D UNet for skull stripping of MRI rodent brains. While these methods have shown notable success, a common limitation is their reliance on 2D MRI data, which inherently lacks spatial relationships present in 3D imaging. This restriction could affect their accuracy and generalizability when applied to fMRI datasets, where spatial context is critical.

Hsu et al. \cite{hsu20213d} proposed a method based on the 3D UNet framework which uses 3D convolutional kernels to predict
segmentations on volumetric patches. The method has been trained and evaluated for brain extraction of 3D volumetric rat brain MRI data. Seo et al. \cite{seo2022unified} proposed a method based on inverse spatial normalization and deep convolutional neural network models for brain extraction of positron emission tomography (PET) in mice. These methods have demonstrated success on structural MRI and PET data but remain challenging for functional MRI (fMRI). They fail to address challenges specific to fMRI, including low-resolution data, small brain regions, and the varying slice sizes in front brain slices. 
The 3D UNet is also applied by Ruan et al. \cite{ruan2022automated} for skull stripping in mouse anatomical and functional MRI analysis. However, its performance is limited when dealing with the small brain regions present in the rostral and caudal slices, often resulting in incomplete or inaccurate brain extraction. This challenge arises due to the inability of the 3D UNet to adequately capture the fine-grained spatial details in these regions, which are crucial for precise skull stripping in preclinical fMRI data. 

Transformers have recently been used in medical imaging to enhance feature extraction, improving segmentation accuracy \cite{shamshad2023transformers}, \cite{xiao2023transformers}, \cite{zhang2025transgraphnet}. Lin et al. \cite{lin2024rs2} applied a 3D UNet framework with Swin transformers \cite{liu2021swin} for rodent anatomical MRI skull stripping. Their model is based on Swin-UNETR \cite{hatamizadeh2021swin} to apply in MRI preclinical pipeline. However, this approach needs refinement for preclinical fMRI, which often suffers from noise and low resolution. Accurate extraction of both global and local features is crucial for this application.  
 
In this paper, we propose a model based on Smart Swin Transformer (SST) developed for 2D medical images by Fu et al. \cite{fu2024sstrans} to address the limitations of the standard Swin Transformer for preclinical fMRI skull stripping. While the mask in the conventional Swin Transformer primarily retains dependencies between closely located pixels, the mask in our approach incorporates valuable distant interactions by considering the functional significance of each slice. This enhancement ensures that our method captures both local and global spatial relationships, which are critical for accurate brain extraction in preclinical fMRI data.
We propose a modified SST to streamline the preprocessing pipeline and improve efficiency in preclinical fMRI studies. Our method is designed to be lightweight, fast, and computationally efficient, ensuring adaptability and robustness across different fMRI studies. By introducing structural improvements, our model reduces resource requirements, making it suitable for deployment in resource-constrained environments. 
The SST is used as the encoder to a 3D dense UNet structure to apply densely connected layers reusing features and improving gradient flow, which makes it highly effective in capturing image details. This architecture propagates information from earlier layers to deeper layers and captures detailed features from the input data. 
Preclinical datasets often exhibit imbalances in slice sizes across different regions of the brain (front, middle, and back), posing a challenge to skull stripping algorithms. To address this issue, we employed a combined loss function integrating Dice loss \cite{milletari2016v} and Focal loss \cite{ross2017focal} functions 
to guide the optimization of parameters without increasing computational cost. This approach prevents the model from solely focusing on pixel accuracy across the entire image, ensuring that boundary pixels and other critical areas are also accurately predicted. It helps to handle the challenges of segmenting irregularly distributed and class-imbalanced related to different slice sizes.
Consequently, the SST-based 3D dense UNet model provides a robust and accurate skull stripping method in preclinical fMRI analysis and improves the network's ability to accurately distinguish between brain and non-brain tissues.
Comparative evaluations demonstrate its state-of-the-art performance and practicality for preclinical applications. The contributions of this work can be summarized as follows.
\begin{itemize}
    \item We develop a novel skull stripping approach for preclinical fMRI using a dense 3D UNet combined with a SST-based feature extractor.
    \item We use a combination of Dice and Focal loss functions to improve accuracy and robustness in noisy, low-resolution fMRI data with different slice sizes. 
    \item We show that the proposed method outperforms existing techniques in skull stripping quality, Dice similarity score, and fMRI downstream tasks.
\end{itemize}

The rest of the paper is organized as follows. Section~\ref{sec 2} proposes the skull stripping algorithm using SST and dense UNet.
Section~\ref{sec 3} provides datasets details information. Section~\ref{sec 4} covers the experiments and evaluation results. Finally, Section~\ref{sec 5} concludes the paper. 

\section{Proposed Method}
\label{sec 2}
\subsection{Smart Swin Transformer}
\begin{figure*}
\begin{center}
\includegraphics[width=\textwidth]{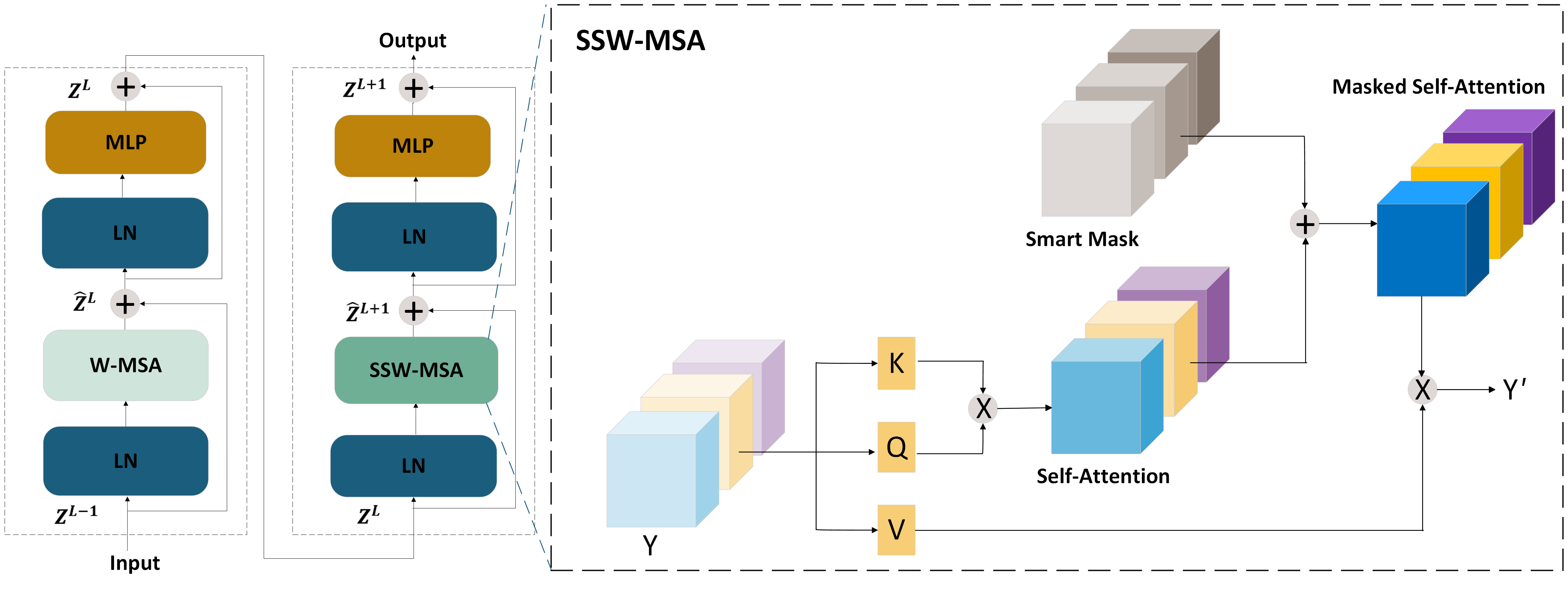}
\end{center}
\caption 
{ \label{1}
On the left side of the figure, the Smart Swin Transformer, composed of a W-MSA and an SSM-MSA, is shown. The right side illustrates the detailed computation process of the SSW-MSA, with the varying colors in the Smart Mask representing the distinct masks used across different channels.} 
\end{figure*}

We apply Smart Swin Transformer (SST) \cite{fu2024sstrans} which is originally proposed for 2D medical image segmentation. The SST is employed in our proposed method to extract features for preclinical fMRI skull stripping as shown in Figure \ref{1}. The Swin Transformer \cite{liu2021swin} consists of two consecutive modules: Multi Self-Attention (W-MSA) and Shifted Window Self-Attention (SW-MSA). 
In the context of 3D MRI data, these modules are extended to capture both spatial and volumetric features across different brain regions. The W-MSA, which primarily focuses on local information, is used to model voxel-wise interactions within a 3D window, ensuring that fine anatomical structures are preserved. The SW-MSA, which captures global interactions, is adapted to process cross-window dependencies across 3D feature volumes.
However, in standard SW-MSA, the masks of the self-attention matrix remain fixed across all channels, limiting the ability to leverage the diverse feature representations required for accurate skull stripping. To address this, the Smart Mask-based SST dynamically determines the mask of each channel based on its specific characteristics. This enables adaptive learning of voxel-wise and regional interactions, effectively distinguishing brain tissues from non-brain structures. By leveraging channel-adaptive attention, the model can enhance feature extraction across different MRI contrasts, improving brain extraction accuracy while preserving anatomical details. 

The primary function of the SSW-MSA (Smart Shifted Window Multi-Head Self-Attention) is illustrated for one slice in Figure \ref{2}. The movement of the blue box represents the shifted window. The orange box denotes the region used for computing the self-attention score, which is constrained by the Smart Mask within a specific channel. Each channel has distinct masks, and only one mask is displayed here (specifically, the channel responsible for capturing interactions among foreground pixels).

\begin{figure}
\begin{center}
\includegraphics[width=10 cm]{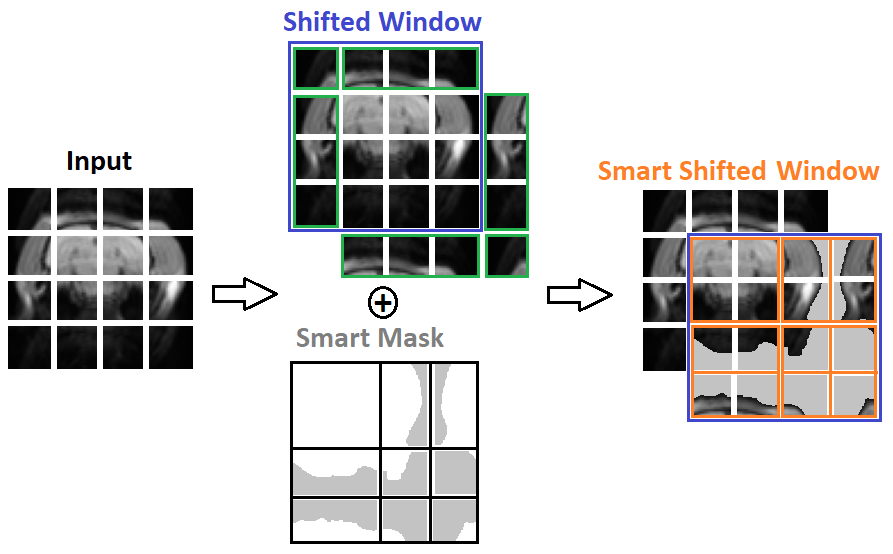}
\end{center}
\caption 
{ \label{2}
The shifted window and Smart Mask within the SSW-MSA (Smart Shifted Window Multi-Head Self-Attention) mechanism of the Smart Swin Transformer (SST), specifically for one slice and one channel.} 
\end{figure}

\begin{figure}
\begin{center}
\includegraphics[width=11 cm]{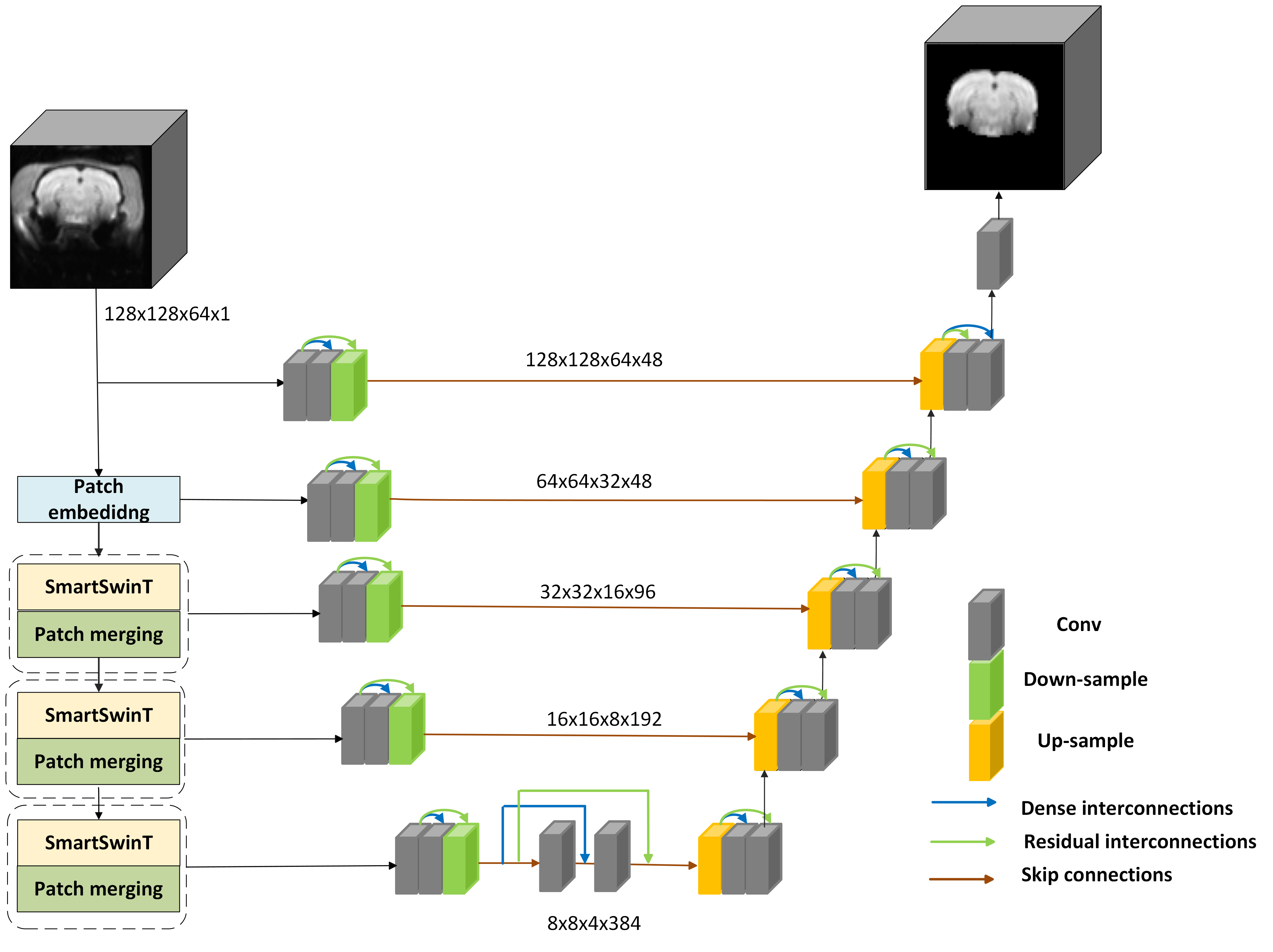}
\end{center}
\caption 
{ \label{3}
A Framework of the proposed model, SST-DUNet.} 
\end{figure}

The self-attention module is defined by the following equation:

\begin{equation}
\label{eq. 1}
\text{Self\_Attention}(Q, K, V) = \text{Softmax}\left(\frac{QK^T}{\sqrt{d}} + \text{SmartMask}\right) V
\end{equation}

where $Q$, $K$, and $V$ are the Query, Key, and Value matrices, respectively, with dimensions $Q, K, V \in \mathbb{R}^{M^3 \times d}$. $M^3$ is the number of patches in the input volume. 
$d$ is the dimension of the key/query vectors.
${SmartMask} \in \mathbb{R}^{M^3 \times M^3}$ is the channel-specific mask that is learned to adjust the self-attention weights for each channel.

The smart mask takes advantage of channel-wise features by creating different masks in various channels. The SSW-MSA improves long range dependencies by keeping precise context data and ignoring pointless interactions.
The details of the SSM-MSA are shown on the right side of Figure \ref{1}. W-MSA and SSW-MSA provide the local and global information focusing on features at channel level which is different from the Swin transformer. The formula to represent the Swin transformer block, the structure on the left side of the Figure \ref{1} is as follows:

\begin{equation}
\begin{aligned}
Z^l &= W_{\text{MSA}}(\text{LN}(\hat{Z}^{l-1})) + \hat{Z}^{l-1} \\
Z^l &= \text{MLP}(\text{LN}(\hat{Z}^l)) + \hat{Z}^l \\
\hat{Z}^{l+1} &= SSW_{\text{MSA}}(\text{LN}(\hat{Z}^l)) + \hat{Z}^l \\
Z^{l+1} &= \text{MLP}(\text{LN}(\hat{Z}^{l+1})) + \hat{Z}^{l+1}
\end{aligned}
\end{equation}

where $Z^i$ is the output of layer $i$. $LN$ (LayerNorm) represents a regularization. $MLP$ denotes multi-layer preceptron. $W_{MSA}$ and $SSW_{MSA}$ represent the window self-attentive mechanism and the smart shifted window self-attentive respectively.

\subsection{Network Architecture}
We apply dense UNet architecture with SST for our proposed SST-DUNet preclinical fMRI skull stripping. Dense UNet was initially developed by Kolavrik et al.\cite{kolavrik2019optimized} for the human brain and spine segmentation. 
In the dense UNet architecture, dense connections are introduced between layers, meaning each layer receives input not only from its preceding layer but also from all previous layers. This dense connectivity allows the model to reuse features from different scales and levels of abstraction, which can be particularly beneficial for handling preclinical data.
Preclinical MRI images may lack fine details, and traditional UNet architectures may struggle to capture and preserve important features for accurate skull stripping. The dense connections in dense UNet facilitate the flow of information throughout the network, enabling better feature reuse and enhancing the model's ability to extract relevant details. Dense UNet architectures are known for their ability to handle both local and global contexts leading to more accurate and robust segmentation results.

The SST-DUNet architecture is shown in Figure \ref{3}. The dense UNet architecture is composed by integrating extra interconnections into the core structure of the 3D UNet architecture. Dense and residual interconnections are shown in Figure \ref{3} in blue and green color, respectively.
As the figure shows an image of  128 × 128 × 64 is input to the SST and encoder part of the dense UNet. The first layer increases the number of channels to 48. There is a patch embedding, three SST blocks, and three patch merging to downsample the features by a factor of two each time. 
These four layers of the SST output features with channel numbers of 48-96-192-384 respectively. 
The features of these four scales are input into the decoder part of the network.
This architecture is characterized by an auto-encoder design featuring five down-sampling and five up-sampling blocks linked by a bridge block at the bottom of the network. 
Each down-sampling block reduces the feature size by half through a max-pooling layer.
At the start of each up-sampling block, the feature size is doubled through a transposed convolution layer employing a stride of 2 for each dimension. Similarly, the decoder module utilizes the same number of layers in reverse order. Each convolutional layer of the encoder
and decoder, except the last one, is followed by a Leaky rectified linear unit (LeakyReLU) activation function. A Sigmoid activation function is applied in the last layer. The Sigmoid function is used to produce output values that represent probabilities, enabling the network to generate binary skull stripping masks.

\subsection{Combo Loss Function}
A key step to train deep networks on imbalanced data is to apply a proper loss function.
Unlike the common approach of combining Cross-Entropy (CE) loss \cite{liu2020survey} and Dice loss \cite{milletari2016v} functions, which primarily target the overall segmentation accuracy of the entire image, the integration of Focal loss \cite{ross2017focal} and Dice loss functions prioritizes challenging-to-classify pixels, such as those at boundaries or across varying scales. This strategy enables the SST to effectively identify and utilize high-quality long-range dependencies across different channels, thereby enhancing the model's segmentation accuracy.  
The combo loss ($L_{Combo}$) is calculated as a weighted sum of  Dice loss ($L_{Dice}$) and Focal loss ($L_{F}$) defined by the following formula.

\begin{equation}
\label{eq. 3}
L_{Combo}=\alpha L_{F}+(1-\alpha)L_{Dice}
\end{equation}

where $\alpha\in[0,1]$ is used to control the relative contribution of the Dice and Focal terms to the loss.

Dice loss is a widely adopted metric to assess the dissimilarity between predicted ($X$)and ground truth ($Y$) segmentation regions:

\begin{equation}
\label{eq. 4}
L_{Dice}=1- \frac{2 \left| X \cap Y \right|}{\left| X \right| + \left| Y \right|}
\end{equation}

Dice loss is minimized when the model maximizes the overlap
between predicted and true segmentation regions. However, there are limitations because of the linear nature of Dice loss.
To address these drawbacks, we apply the Focal loss function which is obtained by adding an exponential weight term to the Cross-Entropy (CE) loss ($L_{CE}$). 
The CE and Focal loss functions are calculated as follow:

\begin{equation}
\label{eq. 5}
\begin{split}
L_{CE} &= -\log(p_t) \\
L_F &= -(1 - p_t)^{\gamma} \log(p_t)
\end{split}
\end{equation}

where, $p_t$ refers to the classification property of the current pixel. $\gamma$ is a hyper-parameter designed to regulate the intensity of penalizing incorrectly classified pixels, typically set to 2. As evident from the equation, Focal loss imposes a stronger penalty on pixels that are more likely to be misclassified. This approach helps address the imbalance between positive and negative samples, as well as the class imbalance.
 
\subsection{Implementation Details}
The model was trained using the Keras framework using TensorFlow backend on the Nvidia GeForce RTX 4090 graphics card using CUDA.  
The model was trained with a combo loss, and AdamW optimizer   with parameters learning rate$=0.00001$, $\alpha=0.4$, and $decay=0.0001$, and batch size of 2. 
The learning rate scheduler uses Linear Warmup Cosine Annealing, starting with a 50-epoch warmup phase where the learning rate slowly rises from a low value to the target rate. After warmup, the learning rate follows a cosine annealing pattern, gradually decreasing to improve convergence.

\section{Datasets}
\label{sec 3}
In this paper, we applied three in-house datasets related to seven studies containing 280 rats collected by the Center for Translational NeuroImaging (CTNI) of Northeastern University.
Experiments were conducted using a Bruker Biospec 7.0T/20-cm USR horizontal magnet (Bruker, Billerica, Massachusetts) and a 20-G/cm magnetic field gradient insert (ID = 12 cm) capable of a 120-µs rise time (Bruker). Radiofrequency signals were sent and received with the quad-coil electronics built into the animal restrainer. Male Sprague Dawley rats weighting 325-350 g were obtained from Charles River Laboratories (Wilmington, MA, USA), and maintained on a 12:12 hour light-dark cycle with a light on at 0700 hours and allowed access to food and
water ad libitum. All rats were acquired and cared for in accordance with the guidelines published in the Guide for the Care and Use of Laboratory Animals (National Institutes
of Health Publications No. 85–23, Revised 1985) and adhered to the National Institutes of Health and the American Association for Laboratory Animal Science guidelines. The protocols used in this study were compliant with the regulations of the Institutional Animal Care and Use Committee at the Northeastern University.

At the beginning of each imaging session, a high-resolution anatomical data set was collected using the rapid acquisition with relaxation enhancement (RARE) pulse sequence (25 slices; 1 mm; field of view (FOV) 3.0 cm ; image size 256 × 256; Time of Repetition (TR) 2.5 sec; Time of Echo (TE) 12.4 msec; Number of Excitations (NEX) 6; 6-minute acquisition time). 
Functional images for the task-based fMRI portion were acquired using a Half-Fourier acquisition, single-shot, turbo-spin echo sequence (RARE-st). A single scanning session was acquired 96×96 in-plane resolution, 20-25 slices every 6 seconds (TR = 6000 msec; TE 48 msec; RARE factor 36; NEX 1) repeated 100 times for every 10-minute scan. 
Resting state functional MRI (rsfMRI) data were collected before and after task-based fMRI scanning using a spin-echo triple-shot EPI sequence (imaging parameters: image size 96 × 96 × slice numbers  [length × width × depth], the number of slices varies between 20 and 25, depending on brain size, TR 1000 ms, TE 15 ms, voxel size 0.312 × 0.312 mm, slice thickness 1.2 mm, 300 repetitions, acquisition time 15 min). 

\subsection{Data Pre-processing}
Data preprocessing is critical for enhancing the model's robustness and generalization. All the images are resized to 128 × 128 × 64. Additionally, data augmentation methods such as Gaussian noise, Gaussian blur, brightness and contrast adjustments, simulated low resolution, and gamma correction were incorporated to enhance dataset diversity and improve the model's ability to handle real-world image variations.
The input voxel values were adjusted to fall within the range of 0.0 to 1.0 through normalization.

\subsection{Data for Training and Test}
In this paper, we applied the mean of fMRI data to create a 3D MRI dataset. We applied 5 studies (dataset 1) containing 260 samples for training and test. To train the model, we initially created a training dataset by randomly selecting 80\% of the data, reserving the remaining 20\% for final performance testing. During the training phase, an additional 80\% of the data were randomly sampled from the training dataset, leaving the remaining 20\% for validating the training of the model. This training-validation process was repeated five times to ensure an unbiased data distribution. Subsequently, the model with the highest averaged validation accuracy was chosen as the final model for testing.
To evaluate the model's generalization ability, we applied two studies, each of them containing 10 samples (dataset 2, and dataset 3). These studies have not been applied for training. 

\subsection{Data Post-processing}
After generating the predicted mask, post-processing involves retaining the largest connected component while discarding smaller, disconnected regions. This ensures a coherent and structurally meaningful skull-stripped output for fMRI data. By eliminating noise and irrelevant fragments, the method enhances segmentation accuracy and preserves anatomical integrity. This approach is particularly beneficial for fMRI preprocessing, where accurate skull removal is essential for downstream analysis. This simple effective strategy refines the network’s output, improving reliability for real-world applications. Ultimately, it ensures that the final mask aligns with the structural requirements of fMRI data processing.

\section{Experimental results}
\label{sec 4}
To validate the effectiveness of our proposed approach, we compared it with RATS \cite{oguz2014rats}, a commonly used traditional method, and deep learning-based methods including 3D UNet \cite{hsu20213d}, UNet-CNN \cite{porter2024fully}, SWIN-UNETR\cite{hatamizadeh2021swin}, SSTrans-Net \cite{fu2024sstrans}, and RS2-Net \cite{lin2024rs2}. RATS is a traditional intensity-based method that relies on region-growing and thresholding techniques. 3D UNet has been studied in preclinical fMRI skull stripping research by Ruan et al. \cite{ruan2022automated} which employs an encoder-decoder structure with skip connections. UNet-CNN applies a modified UNet-based framework for feature extraction. Swin-UNETR introduces a transformer-based approach with a sliding window mechanism for enhanced feature extraction. SSTrans-Net is a 2D medical segmentation method based on a u-shaped Smart Swin Transformer.
RS2-Net is based on Swin-UNETR for preclinical MRI skull stripping. The parameters for each method were selected based on the optimal parameters recommended by each respective paper. 

To quantitatively assess the skull stripping performance of models, we measured the similarity between the resulting brain and manually drawn brain masks, which served as the ground truth. Brain masks were manually computed for each sample with \href{http://www.itksnap.org/pmwiki/pmwiki.php} {ITK-SNAP} (v3.8.0). To improve contrast, mean files of the functional images were used to delineate between the brain and non-brain tissue. 

\subsection{MRI Image Analysis}
In this section, we present the MRI image analysis methods used for skull stripping, focusing on the evaluation of the proposed approach using various quantitative metrics.

\subsubsection{Evaluation Metrics} 
We applied the following evaluation metrics to assess the proposed model's performance quantitatively. 
Dice is used to compute the similarity score of two samples. Positive Prediction Value (PPV) is applied to calculate the proportion of true positives within the predictions made. 
An evaluation of surface distance using the Hausdorff Distance (HD), which measures the distance between two sample sets. And Sensitivity (SEN) calculates the rate of true positives in manual skull stripping. We consider $X$ to represent the voxel set of the manually skull-stripped volume, and $Y$ to represent the voxel set of the predicted volume. 
These metrics are calculated as follows:

\begin{align}
\text{Dice} &= \frac{2 \left| X \cap Y \right|}{\left| X \right| + \left| Y \right|} \\
\text{PPV} &= \frac{\left| X \cap Y \right|}{\left| Y \right|} \\
\text{HD} &= \max \{h(X, Y), h(Y, X)\} \\
\text{SEN} &= \frac{\left| X \cap Y \right|}{\left| X \right|}
\end{align}

where $h(X, Y) = \max \left\{ \min d(x, y) \mid x \in X, y \in Y \right\}
$ and $d(x, y)$ is the Euclidean distance between $x$ and $y$.

\subsubsection{Results}

\begin{table*}
\caption{\textbf{Metrics values are reported as mean $\pm$ standard deviation on CTNI fMRI dataset 1. Statistical analysis is conducted to compare the performance of the proposed SST-DUNet and those of other competing methods $^*P < 0.05$, $^{**}P < 0.01$.}}
\label{1}
\centering  
\begin{tabular}{c c c c c} 
\hline
\rule[-1ex]{0pt}{3.5ex} Method&Dice&PPV&HD&SEN\\
\hline
\rule[-1ex]{0pt}{3.5ex} RATS&0.9032$\pm$0.0317$^{**}$&0.8956$\pm$0.0345$^{**}$&5.245$\pm$2.512$^{**}$&0.9254$\pm$0.0045$^{**}$\\
\rule[-1ex]{0pt}{3.5ex} 3D UNet&0.9648$\pm$0.0221$^{*}$&0.9673$\pm$0.0186$^{**}$&2.619$\pm$0.438$^{*}$&0.9362$\pm$0.0175$^{**}$\\
\rule[-1ex]{0pt}{3.5ex} UNet-CNN&0.9749$\pm$0.0210$^{**}$&0.9781$\pm$0.0134$^{**}$&2.178$\pm$0.424$^{**}$&0.9465$\pm$0.0115$^{**}$\\
\rule[-1ex]{0pt}{3.5ex} Swin-UNETR&0.9785$\pm$0.0128$^{**}$&0.9804$\pm$0.0117&2.109$\pm$0.271$^{**}$&0.9764$\pm$0.0098$^{*}$\\
\rule[-1ex]{0pt}{3.5ex} SSTrans-Net&0.9793$\pm$0.0116$^{**}$&0.9815$\pm$0.0123&1.958$\pm$0.339$^{**}$&0.9752$\pm$0.0114$^{*}$\\
\rule[-1ex]{0pt}{3.5ex} RS$^2$-Net&0.9809$\pm$0.0119$^{*}$&0.9823$\pm$0.0108$^{*}$&1.887$\pm$0.243$^{*}$&0.9849$\pm$0.0091\\
\rule[-1ex]{0pt}{3.5ex} SST-DUNet (ours)&\textbf{0.9865$\pm$0.0112}&\textbf{0.9855$\pm$0.0098}&\textbf{1.597$\pm$0.217}&\textbf{0.9861$\pm$0.0086}\\
\hline
\end{tabular}
\end{table*}

Table \ref{1} shows the mean values and standard deviations of all metrics for the CTNI fMRI dataset 1 test set. The table illustrates that our method achieved the highest values in Dice, PPV, and SEN metrics and the lowest value in HD metrics compared with other methods. Although all the learning-based methods achieved ideal results with Dice $> 0.95$, the proposed method, SST-DUNet, illustrated higher accuracy (Dice $= 0.9865\pm0.0112$) versus those other methods. The low PPV and high SEN from the RATS method shows an overestimation of brain extraction. The lower value for HD ($1.597\pm0.217$) further supports the superior ability of our method to match the ground truth. 

\begin{table*}[ht]
\caption{Metrics values are reported as mean $\pm$ standard deviation on CTNI fMRI dataset 2, and dataset 3. Statistical analysis is conducted to compare the performance of the proposed SST-DUNet and those of other competing methods $^*P < 0.05$, $^{**}P < 0.01$.} 
\label{2}
\begin{center}   
\begin{tabular}{c c c c c} 
\hline
\rule[-1ex]{0pt}{3.5ex} Dataset2&&&&\\
\rule[-1ex]{0pt}{3.5ex} Method&Dice&PPV&HD&SEN\\
\hline
\rule[-1ex]{0pt}{3.5ex} RATS&0.8613$\pm$0.0218$^{**}$&0.7522$\pm$0.4819$^{**}$&5.467$\pm$0.492$^{**}$&0.9402$\pm$0.0271$^{**}$\\
\rule[-1ex]{0pt}{3.5ex} 3D UNet&0.9517$\pm$0.0233$^{*}$&0.9659$\pm$0.0179$^{**}$&2.724$\pm$0.385$^{**}$&0.9149$\pm$0.0138$^{**}$\\
\rule[-1ex]{0pt}{3.5ex} SSTrans-Net&0.9649$\pm$0.0161$^{**}$&0.9681$\pm$0.0127$^{*}$&1.962$\pm$0.411&0.9749$\pm$0.0127$^{*}$\\
\rule[-1ex]{0pt}{3.5ex} RS$^2$-Net&0.9714$\pm$0.0154$^{*}$&0.9729$\pm$0.0123$^{*}$&1.812$\pm$0.251&0.9755$\pm$0.0123\\
\rule[-1ex]{0pt}{3.5ex} SST-DUNet (ours)&\textbf{0.9786$\pm$0.0142}&\textbf{0.9788$\pm$0.0199}&\textbf{1.721$\pm$0.220}&\textbf{0.9763$\pm$0.0114}\\
\hline
\rule[-1ex]{0pt}{3.5ex} Dataset3&&&&\\
\rule[-1ex]{0pt}{3.5ex} Method&Dice&PPV&HD&SEN\\
\hline
\rule[-1ex]{0pt}{3.5ex} RATS&0.8995$\pm$0.0216$^{**}$&0.7529$\pm$0.4733$^{**}$&5.455$\pm$0.492$^{**}$&0.9418$\pm$0.0263$^{**}$\\
\rule[-1ex]{0pt}{3.5ex} 3D UNet&0.9525$\pm$0.0217$^{*}$&0.9661$\pm$0.0188&2.709$\pm$0.384&0.9154$\pm$0.0125$^{**}$\\
\rule[-1ex]{0pt}{3.5ex} SSTrans-Net&0.9685$\pm$0.0155$^{**}$&0.9727$\pm$0.0129$^{*}$&1.958$\pm$0.405$^{*}$&0.9751$\pm$0.0116\\
\rule[-1ex]{0pt}{3.5ex} RS$^2$-Net&0.9735$\pm$0.0149$^{**}$&0.9741$\pm$0.0121&1.769$\pm$0.249$^{**}$&0.9764$\pm$0.0112\\
\rule[-1ex]{0pt}{3.5ex} SST-DUNet (ours)&\textbf{0.9804$\pm$0.0131}&\textbf{0.9819$\pm$0.0109}&\textbf{1.633$\pm$0.215}&\textbf{0.9789$\pm$0.0109}\\
\hline
\end{tabular}
\end{center}
\end{table*} 

To demonstrate the generalizability of our method, a test was conducted on two unseen datasets (dataset 2 and dataset 3). The trained model on dataset 1 has been applied and evaluated using unseen data. Table \ref{2} shows the quantitative results of our method and its comparison with state-of-the-art on the dataset 2 and dataset 3. We reported results related to RATS as a conventional method, and three learning-based methods including 3D UNet which was applied by Ruan et al. \cite{ruan2022automated} for fMRI skull stripping and SSTrans-Net, and RS$^2$-Net with highest dice value based on the results in Table \ref{1}. The mean values and standard deviations of all metrics have been reported in this table. As the results show our method outperformed other methods and produced accurate skull stripping results with Dice $=0.9786\pm0.0142$ and HD $=1.7211\pm0.220$ for dataset 2, and Dice $=0.9804\pm0.0131$ and HD $=1.633\pm0.215$ for dataset 3. Results show that the proposed method is capable of performing skull stripping on preclinical brain images across different datasets without additional training.

\begin{figure*}
\begin{center}
\includegraphics[width=\textwidth]{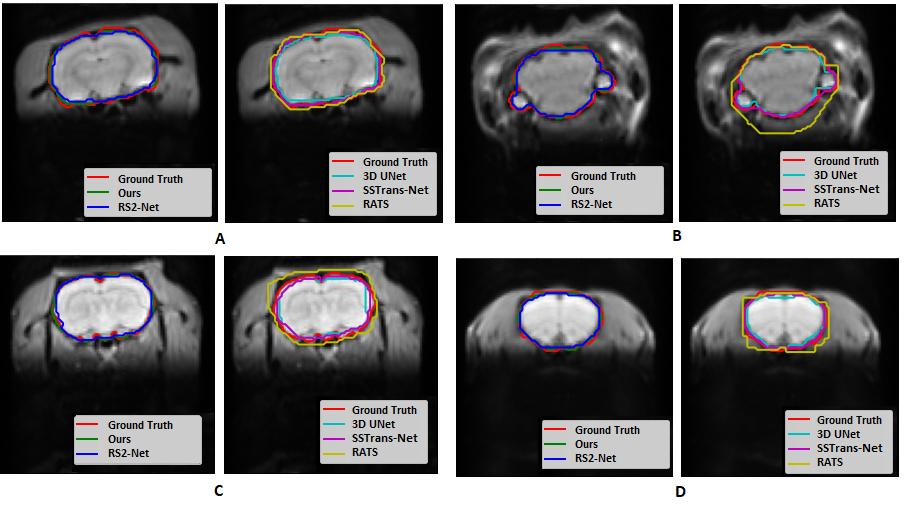}
\end{center}
\caption 
{ \label{4}
This figure presents the skull-stripping results across various CTNI datasets (A, B: dataset 1- C: dataset 2, D: dataset 3). The red region indicates the manually annotated ground truth, while the areas outlined in different colors correspond to the results produced by different methods.} 
\end{figure*}

In addition to the four evaluation metrics discussed in the previous section, we performed a significance analysis on the skull stripping results to assess the differences between the outcomes produced by our method and those from the other methods. The p-values presented in Table \ref{1}, and Table \ref{2} offer important insights into the statistical significance of these differences. A p-value below 0.05 is generally considered statistically significant, meaning the observed differences are unlikely to be due to random chance. In our case, the p-values for most comparisons between our method and other methods, across all  evaluation metrics, are below 0.01, and some of them are below 0.05. This low p-value indicates a substantial and highly significant performance gap, demonstrating that our method outperforms the others in all these metrics.
However, for some metrics in Table \ref{1}, such as PPV and SEN, Swin-UNETR, SSTrans-Net, and RS2-Net show higher p-values. While the metric values still suggest that our method performs better, the differences are not statistically significant. Additionally, in Table \ref{2}, the HD and SEN metrics show higher p-values for SSTrans-Net and RS2-Net on dataset 2,  and the SEN and PPV metrics show higher p-values for SSTrans-Net and RS2-Net on dataset 3.

\begin{figure*}
\begin{center}
\includegraphics[width=10 cm]{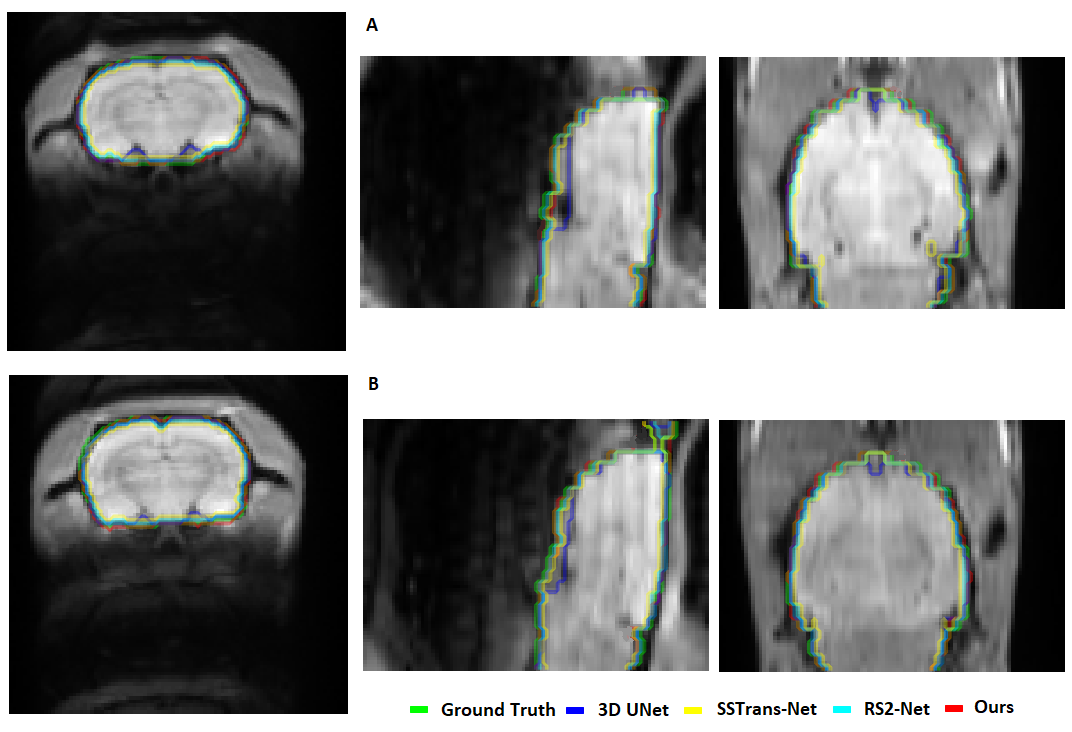}
%\textwidth
\end{center}
\caption 
{ \label{5}
This figure presents the skull stripping results for two fMRI samples from dataset 1 in three differnt views (Axial, Sagittal, and Coronal). The light green region indicates the manually annotated ground truth, while the areas outlined with different colored lines correspond to the outcomes from different methods.} 
\end{figure*} 

To evaluate the performance of our proposed method, skull stripping results from two subjects in the dataset 1 test set are presented in Figure \ref{4}A and B, showing slices from both the middle and back regions of the brain. Additionally, results from a single slice of one subject from dataset 2 and dataset 3 are illustrated in Figure \ref{4}C and D. As shown in Figure \ref{4}, RATS tends to misclassify non-brain tissues as brain, whereas all deep learning-based methods effectively extract brain tissues. Further comparisons of deep learning-based methods applied to two fMRI samples from dataset 1 are visualized in Figure \ref{5}, offering three different views for a more comprehensive assessment.

The Dice metrics for each slice, reported in Table \ref{3}, provide a quantitative comparison of the methods. These results highlight that our proposed method outperforms existing techniques and offers a reliable solution for preclinical skull stripping.

\begin{table}[ht]
\caption{Dice metric values related to Figure \ref{4} and Figure \ref{5}, samples A and B are from dataset 1, sample C, and sample D belongs to dataset 2, and 3 respectively.}
\label{3}
\begin{center}  
\begin{tabular}{c c c c c} 
\hline
\rule[-1ex]{0pt}{3.5ex} Figure \ref{4}&&&&\\
\rule[-1ex]{0pt}{3.5ex} Method&Sample A&Sample B&Sample C&Sample D\\
\hline
\rule[-1ex]{0pt}{3.5ex} RATS&0.9012&0.8756&0.8667&0.8979\\
\rule[-1ex]{0pt}{3.5ex} 3D UNet&0.9536&0.9427&0.9384&0.9415\\
\rule[-1ex]{0pt}{3.5ex} SSTrans-Net&0.9712&0.9678&0.9542&0.9654\\
\rule[-1ex]{0pt}{3.5ex} RS$^2$-Net&0.9751&0.9697&0.9567&0.9686\\
\rule[-1ex]{0pt}{3.5ex} SST-DUNet(ours)&\textbf{0.9819}&\textbf{0.9781}&\textbf{0.9687}&\textbf{0.9773}\\
\hline
\rule[-1ex]{0pt}{3.5ex} Figure \ref{5}&&&&\\
\rule[-1ex]{0pt}{3.5ex} Method&Sample A&&Sample B&\\
\rule[-1ex]{0pt}{3.5ex} 3D UNet&0.9519&&0.9504&\\
\rule[-1ex]{0pt}{3.5ex} SSTrans-Net&0.9642&&0.9617&\\
\rule[-1ex]{0pt}{3.5ex} RS$^2$-Net&0.9661&&0.9654&\\
\rule[-1ex]{0pt}{3.5ex} SST-DUNet(ours)&\textbf{0.9765}&&\textbf{0.9749}&\\
\hline
\end{tabular}
\end{center}
\end{table}

Since fMRI data is inherently noisy, we conducted a test to simulate noise by adding Rician noise. This test aims to assess the robustness of our skull stripping method under varying noise conditions and ensure its effectiveness in real-world scenarios. To further enhance the quality of fMRI data, advanced denoising techniques can be applied prior to skull stripping. In our recent work on fMRI denoising \cite{soltanpour20253d}, we demonstrated an effective approach to reducing noise while preserving essential brain structures. Integrating such methods could further improve skull stripping accuracy, especially in highly noisy datasets. We evaluated the performance of our skull stripping algorithm on images with different noise levels to validate its reliability. 
In this test, we manually introduced Rician Noise \cite{ran2019denoising}, which is an artificially generated noise to the MRI data.
Rician noise is commonly employed in MRI research due to its ability to closely mimic the noise patterns observed in MRI data \cite{ran2019denoising}. As real preclinical data typically exhibits noise levels around $3\%$, we followed the experimental set-up inspired by Ran et al. \cite{ran2019denoising}, applying noise levels (noise standard deviation) ranging from $1\%$ to $15\%$ in increments of $2\%$.

\begin{figure*}
\begin{center}
\includegraphics[width=\textwidth]{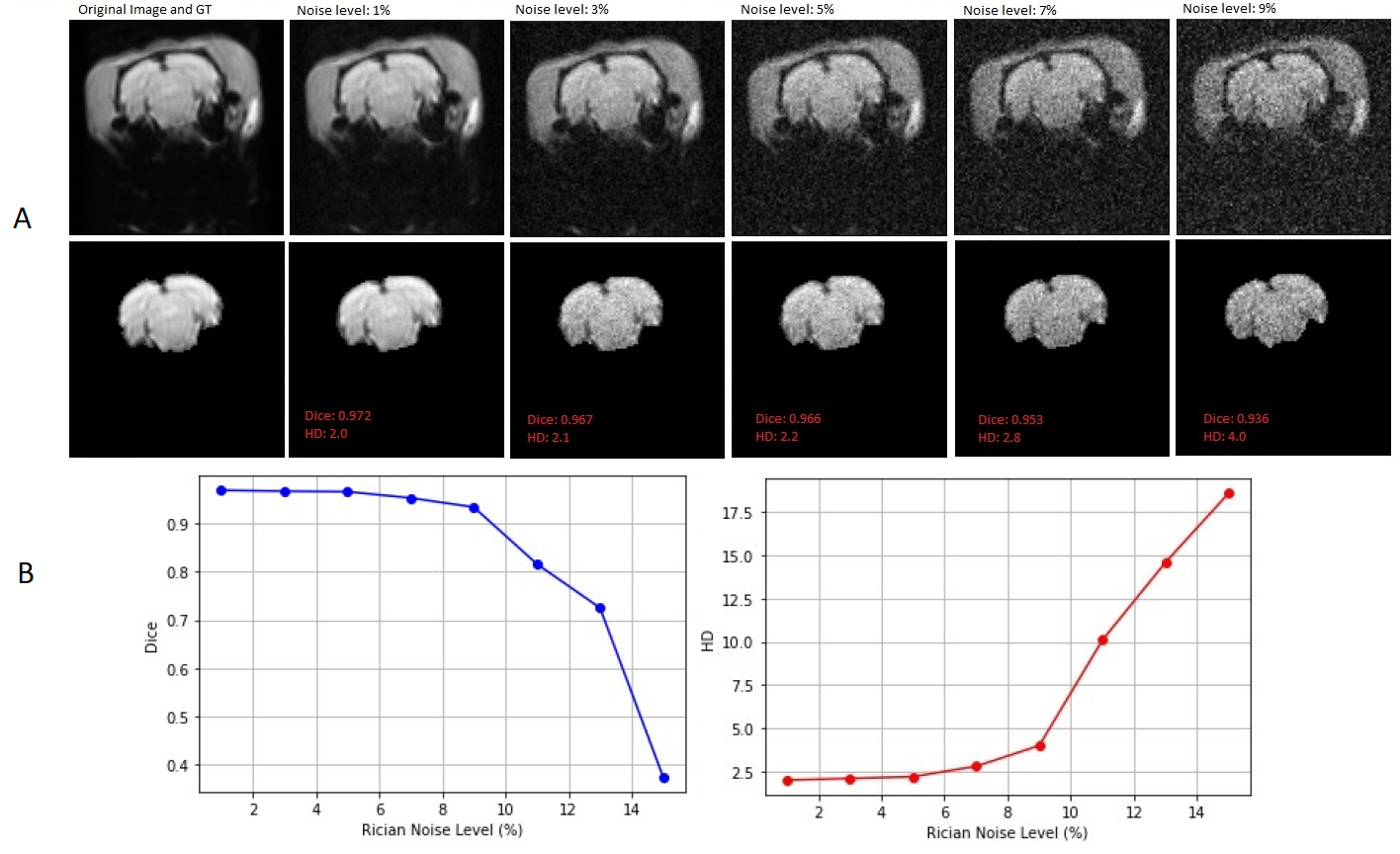}
\end{center}
\caption 
{ \label{6}
Proposed method's performance under different noise levels on dataset 1.} 
\end{figure*}

The model performance has been illustrated in Figure \ref{6} for different noise levels on the dataset 1. Figure  \ref{6}A shows skull stripping results for noise levels $1\%$ to $9\%$ for one sample from the dataset 1 test set for a middle slice. As Figure \ref{6}B shows, our method's performance decreases with increasing Rician noise level. However, the skull stripping accuracy could still reach a suitable Dice $>0.95$ and HD $<3$ for noise levels $1\%$ to $7\%$. 
This illustrates the ability of our method to tolerate noise, implying that this method could effectively manage fMRI data across a broad spectrum of Signal-to-Noise Ratio (SNR) qualities. 

\subsection{Seed-based functional connectivity analysis}
To further compare the manual brain extraction and the proposed SST-DUNet, we applied seed-based functional connectivity analysis \cite{wang2019thalamic}, and \cite{ghaw2024dose}. 
Skull stripping is often a bottleneck in preprocessing for seed-based functional connectivity (FC) analysis due to the time-intensive process of generating brain extraction masks for each subject. Functional connectivity analysis identifies how regions of interest (ROIs) cluster together in the brain, revealing general patterns of connectivity across a group of subjects. To evaluate whether the SST-DUNet algorithm can reliably replace manual skull stripping in FC analysis, we compared the connectivity results from both manual and SST-DUNet methods. The analysis was conducted across five study groups of dataset 1, each comprising eight rat subjects, resulting in $171 \times 171$ matrices of t-stat values representing ROI-to-ROI connectivity. These t-stat values were derived by first calculating Fisher Z-transformed correlation coefficients across the 8 subjects, followed by one-sided t-tests to obtain the final t-statistics.

\begin{figure*}
\begin{center}
\includegraphics[width=\textwidth]{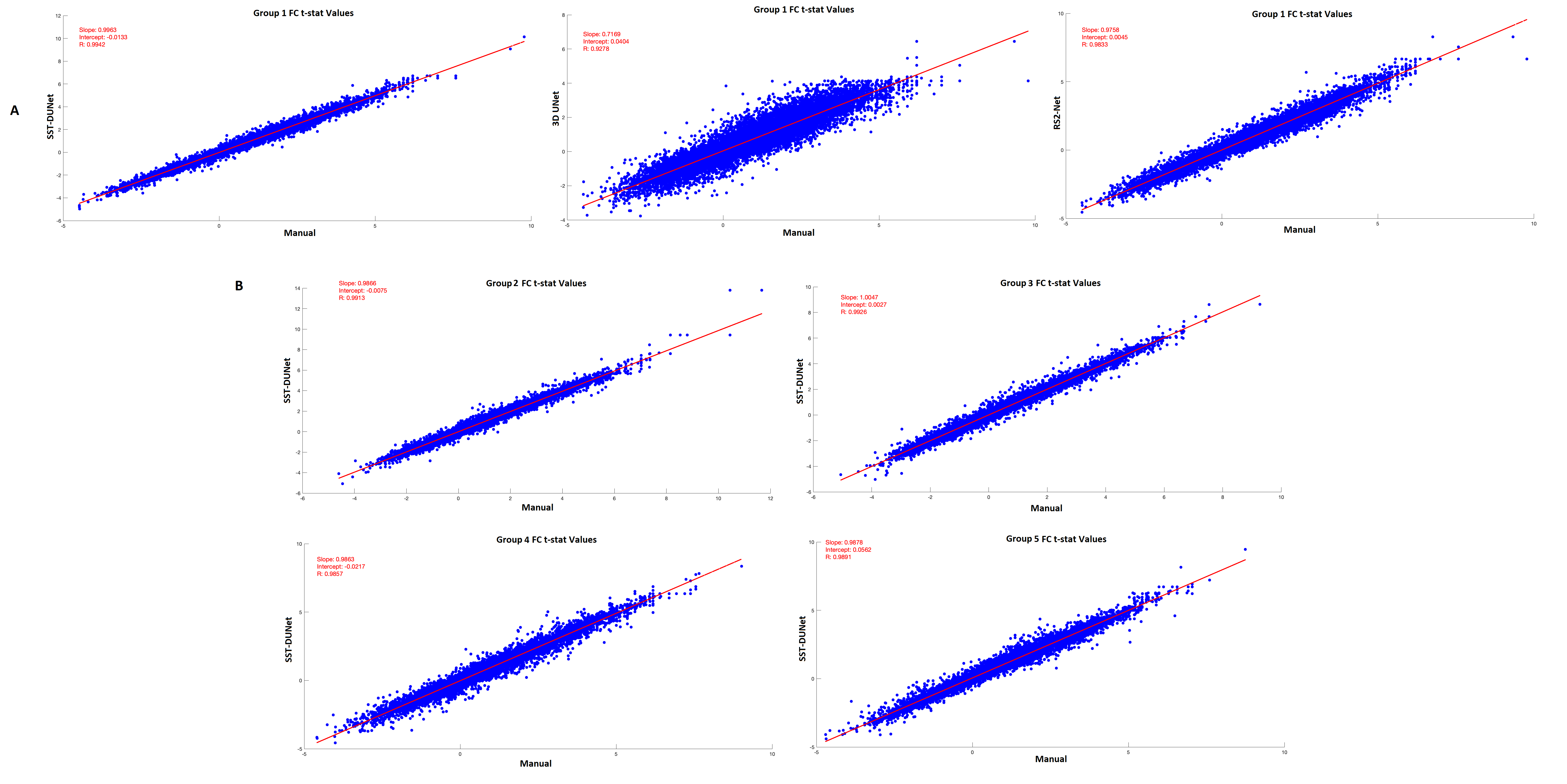}
\end{center}
\caption 
{ \label{7}
FC analysis (A) Analysis of SST-DUNet (this study), 3D UNet, and RS$^2$-Net on group 1 dataset 1. (B) Analysis of SST-DUNet (this study) on Groups 2, 3, 4, and 5 datasets 1.} 
\end{figure*}

A scatterplot for group 1 in Figure \ref{7}A of the t-stat results for manual versus SST-DUNet skull stripping methods showed a strong linear relationship, with a slope of 0.996, an intercept of -0.013, and a Pearson correlation coefficient (R) of 0.994, demonstrating near-identical connectivity outcomes. Also, the SST-DUNet outperforms 3D UNet and RS$^2$-Net as the results in Figure \ref{7}A show. Furthermore, this comparison was repeated in four additional groups with consistent results (see Figure \ref{7}B). The correlation coefficients for those were 0.991, 0.993, 0.986, and 0.989 respectively, providing strong evidence that SST-DUNet provides a reliable and efficient alternative to manual skull stripping for functional connectivity analysis. 

\subsection{Independent Component Analysis}
To further analyze the proposed method for fMRI skull stripping, we applied Independent Component Analysis (ICA).

\subsubsection{Method}
Hypothesis free ICA was executed by combining skull stripping method groups of dataset 1 (Manual=8, SST-DUNet=8, 3D UNet=8, RS$^2$-Net=8) applying the Group ICA of a well-recognized fMRI Toolbox (GIFT, v1.3i, \url{www.nitrc.org/projects/gift/}) in MATLAB \cite{nasseef2015measuring}, \cite{nasseef2019oxycodone}, \cite{nasseef2021chronic}, \cite{charbogne2017mu}, \cite{lupinsky2025resting}. Initially, a 30-component ICA decomposition was performed to identify whole-brain resting-state networks and individual brain regions \cite{nasseef2015measuring}. Additionally, the standard built-in approach within the GIFT toolbox was employed to assess the stability and reliability of the extracted components \cite{nasseef2019oxycodone}.

For computation and statistical analysis, an in-house MATLAB script was utilized from our previous studies \cite{nasseef2019oxycodone},\cite{nasseef2021chronic}. Particularly, the toolbox-generated time series for each of the 30 components were extracted from all 32 subjects (Manual=8, SST-DUNet=8, 3D UNet=8, RS$^2$-Net=8) for functional connectivity (FC) analysis. Moreover, statistical testing included applying two-tailed paired and one-sample t-tests independently with false discovery rate (FDR) correction (Using \texttt{fdr\_bh} function in MATLAB), was conducted to assess component-to-component connectivity.

For each of the 30 components, the correlation coefficient (CC) was computed across all 32 subjects. Consequently, the mean CC and standard deviation (Std) were then calculated for each different preprocessing groups separately. Subsequently, within-group significance testing was performed using a one-sample t-test $(p < 0.05)$. Additionally, a pairwise two-tailed t-test $(p = 0.05)$ was conducted to compare between the Manual and three automated skull stripping methods. Moreover, all p-values were corrected for multiple comparisons using the false discovery rate (FDR) method \cite{nasseef2019oxycodone}, \cite{nasseef2021chronic}.

Furthermore, to visualize group-level outcomes, scatter, box, beeswamp and histogram plots were produced using built-in MATLAB functions. Importantly, these visualization techniques provide a clear representation of potential correlations, data distribution, and group differences, improving the interpretability of our findings.

\subsubsection{Results}
Spatial independent component analysis (ICA) was conducted on the preprocessed resting-state fMRI (rs-fMRI) data from all 32 rats across all the Manual and three automated skull stripping methods. 

\begin{figure*}
\begin{center}
\includegraphics[height=14 cm]{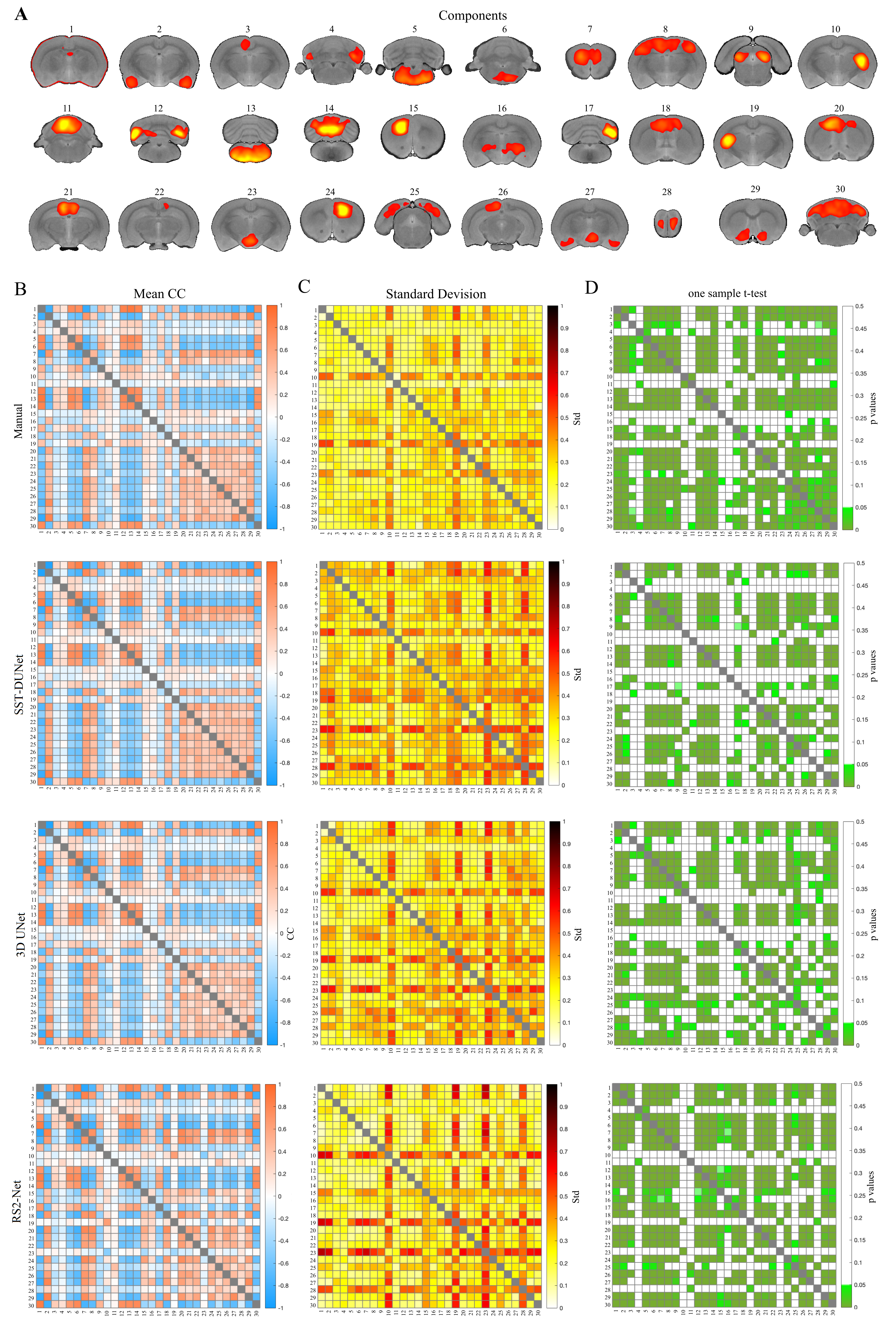}
\end{center}
\caption 
{ \label{8}
ICA 30-Component and Intra-Group Post-Analysis in Manual and other 3 automated skull stripping methods on Rats fMRI data preprocessing; (A) 30-Components (B) Mean Correlation Coefficient (CC) Matrices (C) Standard Deviation (Std) Matrices (D) Statistical Significance Testing (B-D) The right-side color scales for each panel correspond to the magnitude of mean CC values (left column), standard deviation values (middle column), and statistically significant p-values (bottom column).} 
\end{figure*}

Figure \ref{8} has been shown in four parts as follows: (A) 30-Components: Representative 2D illustrations of 30 generated components derived from group independent component analysis (ICA) from all 32 rats (Manual=8, SST-DUNet=8, 3D UNet=8, RS$^2$-Net=8)  fMRI data, highlighting distinct/multiple brain regions or networks across the entire brain. These components are displayed using a red-to-yellow color scale, overlaid on coronal slices of the standard rat brain template, providing a comprehensive visualization of spatial patterns observed in the dataset. Warmer colors (yellow) indicate higher intensity brain regions within each component; (B) Mean Correlation Coefficient (CC) Matrices: The four rows (top to bottom) represent the mean CC square matrices for the Manual group and other three individual automated skull stripping methods independently computed across all 30 independent components. Positive and negative mean CC values are represented using red and blue color bars, respectively; (C) Standard Deviation (Std) Matrices: The corresponding standard deviation matrices for the Manual and other three individual automated skull stripping methods are shown in the four rows. The color scale (ranging from white to dark red) represents the variability in CC values across subjects within each group, providing insight into intra-group consistency; (D) Statistical Significance Testing: The results of the one-sample t-test $(p < 0.05$, FDR-corrected, $n = 8)$ are displayed, where statistically significant p values are indicated by green-shaded bars. (B-D) The right-side color scales for each panel correspond to the magnitude of mean CC values (left column), standard deviation values (middle column), and statistically significant p-values (bottom column).

Using the GIFT toolbox in MATLAB \cite{nasseef2015measuring}, \cite{nasseef2019oxycodone}, \cite{nasseef2021chronic}, \cite{charbogne2017mu}, \cite{lupinsky2025resting}, we first extracted 30 independent components (Figure \ref{8}A) and their time series from each dataset. Subsequently, the quality, stability, and reliability of these components were assessed using the standard built-in validation methods of the GIFT toolbox \cite{nasseef2019oxycodone}, \cite{nasseef2021chronic}. This hypothesis-free ICA approach enabled a whole-brain investigation of potential differences between two Manual and automated skull stripping methods.

To evaluate intra-group connectivity patterns, we computed symmetric correlation coefficients (CC) among the 30 components for each rat individually. Subsequently, for each group, we derived the mean CC (Figure \ref{8}B), standard deviation (Std) (Figure \ref{8}C), and one-sample t-test $(p < 0.05$, FDR-corrected, $n = 8)$ (Figure \ref{8}D) results. Interestingly, these analyses revealed a high degree of similarity in functional connectivity patterns between the Manual and automated skull stripping methods. 

\begin{figure*}
\begin{center}
\includegraphics[width=\textwidth]{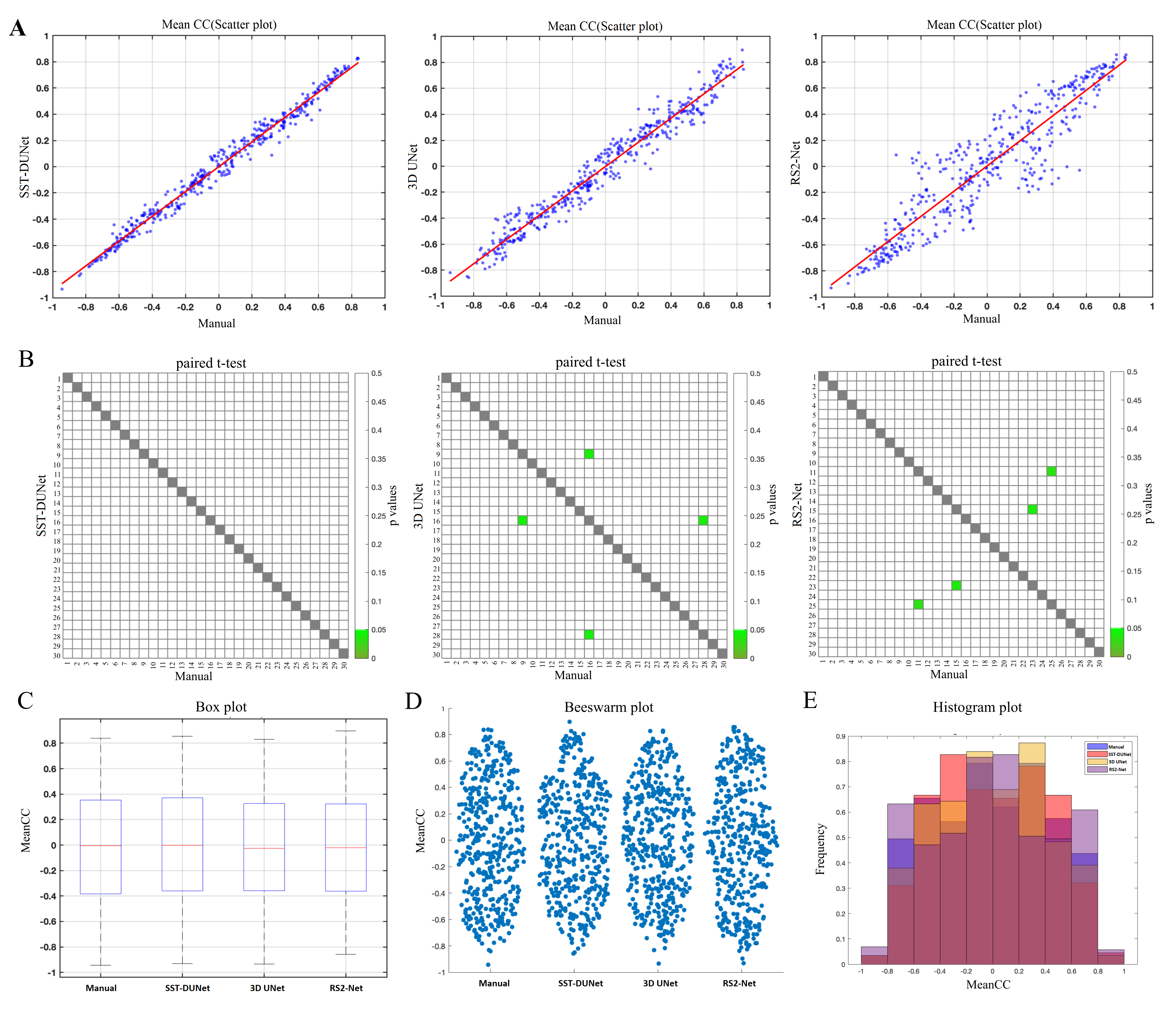}
\end{center}
\caption 
{ \label{9}
Inter-Group ICA 30-Component Post-Analysis in Manual and other three individual automated skull stripping Groups in Rats; (A) Scatter Plot with Regression Line (B) Statistical Significance Testing (C) Box Plot Representation(D) Beeswamp Plot Representation (E) Histogram Plot Representation.} 
\end{figure*}

To further evaluate group-level differences, we illustrate Inter-Group ICA 30-Component Post-Analysis in Manual and other three individual automated skull stripping Groups in Rats in Figure \ref{9}. This figure is presented as follows: (A) Scatter Plot with Regression Line: The left, middle, and right panels illustrate the comparison between the Manual and other three individual automated skull stripping methods independently using scatter plots with fitted regression lines in red color. These plots represent the distribution of 435 correlation coefficient (CC) values across the different skull stripping methods; (B) Statistical Significance Testing: The left, middle, and right panels present the results of a pairwise two-tailed t-test ($p = 0.05$, FDR-corrected, $n = 8$ per group), evaluating statistical differences between the Manual and other three individual automated skull stripping methods distinctly across all 435 CC values; (C) Box Plot Representation: The box plot illustrates the distribution of mean CC values across all 435 samples for the Manual and three individual automated skull stripping methods. It highlights the variation, central tendencies, and spread of the data; (D) Beeswamp Plot Representation: The beeswarm plot provides a detailed visualization of the CC distributions for the Manual and three automated skull stripping methods, capturing individual data points while showcasing variation across all 435 CC values;  (E) Histogram Plot Representation: The histogram depicts the frequency distribution of CC values across 10 bins for the Manual and three automated skull stripping methods, offering a comparative view of their statistical distributions.

We visualized the data using scatter plots with regression lines (Figure \ref{9}A), box plots (Figure \ref{9}C), beeswamp plots (Figure \ref{9}D) and histogram plots (Figure \ref{9}E). Interestingly, the box plot represents a highly similar mean, quartile and outlier between the Manual and SST-DUNet methods. Additionally, a pairwise two-sample t-test $(p = 0.05$, two-tailed, FDR-corrected, $n = 8/8)$ was conducted to compare the Manual and 3 automated skull stripping methods (Figure \ref{9}B). Remarkably, no statistically significant differences were observed between the Manual and proposed SST-DUNet methods, highlighting a striking similarity in their connectivity signatures. Hence, these findings indicate that the proposed method reliably reproduces results obtained through manual processing, underscoring its robustness and potential for automated rs-fMRI analysis.

\subsection{Computational cost}
To assess the model's performance in terms of computational cost, we compared it with different methods. All the experiments have been performed using an RTX 4090 GPU; the results are shown in Table \ref{4}. As the table shows the average inference time reduces from 0.62 s for RS$^2$-Net to 0.55 s for our method. This advantage is important for running models on devices with limited GPU power, making the method more practical for various applications. It also confirms the strong performance of our method in deep learning-based preclinical skull stripping.

\begin{table}[ht]
\caption{Performance comparison. The inference time is measured as the average runtime across all successfully obtained results in the test examples.} 
\label{4}
\begin{center}   
\hfill    
\begin{tabular}{c c c c c c} 
\hline
\rule[-1ex]{0pt}{3.5ex} Methods&Input size&Parameters (M)&Inference time (s)\\
\hline
\rule[-1ex]{0pt}{3.5ex} RATS&Original size&-&179.22\\
\rule[-1ex]{0pt}{3.5ex} 3D UNet&128x128x64&1.9&0.22\\
\rule[-1ex]{0pt}{3.5ex} Swin-UNETR&128x128x64&62.2&0.75\\
\rule[-1ex]{0pt}{3.5ex} SSTrans-Net&128x128x64&29.6&0.68\\
\rule[-1ex]{0pt}{3.5ex} RS$^2$-Net&128x128x64&14.9&0.62\\
\rule[-1ex]{0pt}{3.5ex} Ours&128x128x64&12.5&0.55\\
\hline
\end{tabular}
\end{center}
\end{table}

\subsection{Discussion}
To the best of our knowledge, this study represents the first attempt to perform rat brain fMRI skull stripping using a framework built on the Smart Swin Transformer.
It attains state-of-the-art performance, achieving a Dice coefficient 0.98, comparable to manually segmented brains. Given its outstanding accuracy in skull stripping tasks and reduced computational demands, our framework shows significant potential as an alternative to manual labelling in the rat functional magnetic resonance imaging preprocessing pipeline.
We compared deep learning-based methods with a well-known traditional approach, RATS, and found that they consistently outperform it.
Among the deep learning-based methods, our proposed approach achieved the highest performance, leveraging the strengths of the Smart Swin Transformer (SST) and dense interconnections. This combination enhances feature extraction and information flow, leading to more accurate and consistent skull stripping results across various datasets.

The results of the seed-based functional connectivity analysis demonstrate that SST-DUNet is a reliable and efficient alternative to manual skull stripping for rat brain fMRI preprocessing. The strong linear correlation between manual and SST-DUNet results, with Pearson coefficients consistently above 0.98 across all groups, confirms that the proposed method preserves essential connectivity patterns.

Independent Component Analysis (ICA) offers a powerful, unbiased and hypothesis-free approach for whole-brain analysis, enabling the identification of functionally independent brain regions/networks and facilitating a deeper understanding of brain function and connectivity \cite{nasseef2015measuring}, \cite{nasseef2019oxycodone}, \cite{nasseef2021chronic}, \cite{charbogne2017mu}, \cite{lupinsky2025resting}. More importantly, by decomposing complex neuroimaging data into spatially distinct components, ICA provides valuable insights into intrinsic brain organization, making it a widely used technique in functional connectivity as well as other studies. In this paper, we utilize the advantage of ICA to explore the effectiveness of our newly proposed AI automated skull stripping method in the brain fMRI preprocessing pipeline. Although the scatter plots with regression lines (Figure \ref{9}A), box plot (Figure \ref{9}C) and two-tailed paired t-test (Figure \ref{9}B) indicate high similarity between the Manual and SST-DUNet methods compared to 3D UNet and RS$^2$-Net methods, subtle differences emerge when analyzing the mean CC using one-sample t-test $(p < 0.05$, FDR-corrected, $n = 8)$ (Figure \ref{8}D), beeswarm (Figure \ref{9}D) and histogram (Figure \ref{9}E) plots. Notably, these alternative visualizations reveal trivial discrepancies in the mean CC distributions, suggesting minor variations in how each method processes skull stripping. Moreover, the beeswarm plot highlights individual data point dispersion, while the histogram demonstrates differences in frequency distribution across bins. Hence, these findings underscore the importance of complementary statistical and visualization techniques in assessing preprocessing methods, particularly in studies involving ICA-derived functional connectivity analysis.

While the proposed SST-DUNet framework demonstrates promising results in fMRI skull stripping, there are limitations to consider. The model was trained and evaluated on rat brain datasets, which may affect its performance when applied to other species or brain types. Also, the segmentation accuracy of the developed SST-DUNet model on functional images remains lower compared to anatomical images, primarily due to the lower quality of functional images. Incorporating cross-modality information between anatomical and functional images could potentially enhance the accuracy of skull stripping on functional images.

\section{Conclusion and Future Work}
\label{sec 5}
In this paper, we introduced a novel approach for automatic preclinical fMRI skull stripping, integrating a smart Swin Transformer, 3D Dense UNet, and combo loss. Experimental results demonstrate the superiority of our method compared to both learning-based and conventional approaches. Our evaluation across multiple datasets highlights its reliability, robustness, and generalizability in preclinical rat fMRI analysis.

While our proposed skull stripping method demonstrates strong generalizability, its performance may be influenced by variations in imaging protocols, scanner types, and dataset characteristics. Fine-tuning or retraining may be required to achieve optimal results on highly diverse datasets. Additionally, all our experiments were conducted on rat fMRI data, and further validation on mouse datasets is needed to confirm the method’s adaptability across different preclinical models. Future work will focus on expanding the dataset diversity and evaluating the approach across various imaging conditions to enhance its robustness and applicability.

\bibliographystyle{elsarticle-num-names} 
\bibliography{report}

\end{document}